\documentclass[aps,prb,superscriptaddress,twocolumn,amsmath,amssymb,showpacs]{revtex4}
\usepackage{graphicx}
\usepackage{upgreek}

\begin{document}

\title{Spin diffusion in the Mn$^{2+}$ ion system of II-VI diluted magnetic semiconductor heterostructures}

\author{A. A.~Maksimov}
\email[Corresponding author: ]{maksimov@issp.ac.ru}
\affiliation{Institute of Solid State Physics, Russian Academy of
Sciences, 142432 Chernogolovka, Russia}

\author{D. R.~Yakovlev}
\affiliation{Experimentelle Physik 2, Technische Universit\"at
Dortmund, 44227 Dortmund, Germany} \affiliation{A. F. Ioffe
Physical-Technical Institute, Russian Academy of Sciences, 194021
St. Petersburg, Russia}

\author{J. Debus}
\affiliation{Experimentelle Physik 2, Technische Universit\"at
Dortmund, 44227 Dortmund, Germany}

\author{I. I.~Tartakovskii}
\affiliation{Institute of Solid State Physics, Russian Academy of
Sciences, 142432 Chernogolovka, Russia}

\author{A. Waag}
\affiliation{Institute of Semiconductor Technology, Braunschweig
Technical University, 38106 Braunschweig, Germany}

\author{G. Karczewski}
\author{T. Wojtowicz}
\author{J. Kossut}
\affiliation{Institute of Physics, Polish Academy of Sciences,
02668 Warsaw, Poland}

\author{M.~Bayer}
\affiliation{Experimentelle Physik 2, Technische Universit\"at
Dortmund, 44227 Dortmund, Germany}


\begin{abstract}
The magnetization dynamics in diluted magnetic semiconductor heterostructures based on (Zn,Mn)Se and (Cd,Mn)Te has been studied experimentally by optical methods and simulated numerically. In the samples with nonhomogeneous magnetic ion distribution this dynamics is contributed by spin-lattice relaxation and spin diffusion in the Mn spin system. The spin diffusion coefficient of $7\cdot 10^{-8}~\textrm{cm}^2/\textrm{s}$  has been evaluated for Zn$_{0.99}$Mn$_{0.01}$Se from comparison of experimental and numerical results. Calculations of the giant Zeeman splitting of the exciton states and the magnetization dynamics in the ordered alloys and parabolic quantum wells fabricated by the digital growth technique show perfect agreement with the experimental data. In both structure types the spin diffusion has an essential contribution to the magnetization dynamics.

\end{abstract}


\maketitle

\section{Introduction}

Heterostructures based on II-VI diluted magnetic semiconductors (DMS), like Cd$_{1-x}$Mn$_x$Te or Zn$_{1-x}$Mn$_x$Se, are treated nowadays as model structures for spintronic devices~\cite{Spintronics, Spinbook}.  A large variety of optical spectroscopy techniques has been conveniently used for investigation of their properties due to the strong optical transitions in the vicinity of the band gap being visible in, e.g., absorption, reflection and emission.  DMS materials are known for giant magneto-optical effects originating from the strong exchange interaction of free carriers with the localized magnetic moments of Mn$^{2+}$ ions~\cite{Furdyna88, Die94}.

The spin dynamics of the free carriers in DMS is controlled by the spin and energy transfer in the coupled system of the carriers, magnetic ions and phonon bath (lattice)~\cite{Kon00,Kneip06a,Kneip06b}. The fast exchange scattering of the carriers on the magnetic ions ($10^{-12}-10^{-11}$ s) provides an efficient transfer into the Mn system, but the further transfer from the Mn spin system to the phonon bath is decelerated by the relatively slow spin-lattice relaxation (SLR) rate in materials with low Mn concentrations of a few percent. The control of the dynamical properties of the Mn spins, described by the SLR time, is essential because they may turn out to be a bottleneck for high-speed applications. Therefore, fundamental knowledge about the dynamical properties of DMS and particularly about the dynamics of the Mn spin system is of importance.

In II-VI DMS the SLR time of the localized Mn spins, $\uptau_{\textrm{SLR}}$, is controlled by concentration dependent exchange interactions between the Mn$^{2+}$ ions. It varies by several orders of magnitude from milliseconds down to nanoseconds with increasing Mn concentration $x$ from 0.004 to 0.11, see Fig.~\ref{f1}(c)~\cite{Aki06a,Kneip06b}. Such a strong dependence is explained by the role of Mn clusters as relaxation centers~\cite{Sca96} in the absence of SLR for isolated Mn ions. With increasing Mn concentration the number of clusters grows and the cluster structure evolutes in size involving more and more spins. This shortens a characteristic length for spin transfer from a single Mn ion to a cluster and also accelerates the spin relaxation processes inside the clusters. It has been shown that the digital growth techniques of ordered DMS alloys allow for the same Mn concentration a redistribution of cluster statistics, thus increasing the role of the Mn spin clusters~\cite{Kneip06c}. The cluster formation results in a considerable enhancement of the SLR confirming the leading role of clusters in the magnetization relaxation dynamics.

There are some experimental evidences that spin diffusion in the Mn spin system plays an important role in the magnetization dynamics of II-VI DMS. Firstly, the strong dependence of the SLR time on the Mn concentration can be explained by a nonlocal model of the SLR in DMS where the spin diffusion from single Mn$^{2+}$ ions to Mn clusters provides fast SLR\cite{Suris}. Secondly, direct experimental proof of the spin diffusion contribution to the magnetization relaxation in heteromagnetic DMS structures has been reported for Zn$_{0.99}$Mn$_{0.01}$Se/Be$_{0.93}$Mn$_{0.07}$Te structures with different Mn concentration in the adjacent layers~\cite{Scherbakov05}. The slow SLR channel in the Zn$_{0.99}$Mn$_{0.01}$Se layer has been bypassed by the spin diffusion through the Mn spin system to the Be$_{0.93}$Mn$_{0.07}$Te layer with much faster relaxation rate. Neither experimental nor theoretical reports allowing a quantitative evaluation of the spin diffusion parameters in DMS have been published, yet.

The goal of this paper is to provide a detailed experimental study of the spin diffusion in the Mn spin system of II-VI DMS and to evaluate the spin diffusion coefficient. We took advantages offered by Zn$_{1-x}$Mn$_x$Se/Be$_{1-y}$Mn$_y$Te heteromagnetic structures for a clear experimental emphasis of the spin diffusion contribution to magnetization relaxation dynamics. We developed a numerical approach to describe the magnetization dynamics in structures with a modulated profile of Mn concentration and approved its validity for CdTe/Cd$_{1-x}$Mn$_x$Te digital superlattices and parabolic quantum wells.

\section{Experimentals}

Time-resolved magneto-optical studies of the magnetization dynamics have been performed for various Zn$_{1-x}$Mn$_x$Se and Cd$_{1-x}$Mn$_x$Te based heterostructures grown by molecular-beam epitaxy on (100)-oriented GaAs substrates. Details of the structure design are given in the corresponding Sections in combination with the presentation of experimental data. In this paper we include results for: (i) three Zn$_{1-x}$Mn$_x$Se/Be$_{1-y}$Mn$_y$Te multiple quantum well (MQW) structures with a type-II band alignment, (ii) three CdTe/Cd$_{1-x}$Mn$_x$Te digital superlattices (SL), and (iii)  two CdTe/Cd$_{1-x}$Mn$_x$Te parabolic quantum wells (QW). The CdTe/Cd$_{1-x}$Mn$_x$Te structures have been fabricated by the digital growth technique~\cite{Woj95}.

We used an all-optical experimental technique developed to measure the magnetization dynamics~\cite{Kneip06b,Kneip06a}. It exploits an internal thermometer of the Mn spin tempera\-ture $T_{\textrm{Mn}}$, which is provided by the high sensitivity of the giant Zeeman splitting of excitons (band states) to the polarization of the magnetic Mn ions in an external magnetic field. Taking into account that the changes in the spectral position of the photoluminescence (PL) line $\Delta E_{\text {PL}}(B)$ are proportional to the magnetization changes $\Delta M(B)$, we obtain by means of a magneto-optical technique direct information about magnetization dynamics in DMS heterostructures. For that purpose the studied DMS structures have been exposed to an external magnetic field, a pulsed laser has been used for heating the Mn spin system and the dynamical shift of the PL line has been measured by a time-resolved technique.

The relaxation process of the heated Mn spin system to the bath temperature is described by the magnetization relaxation time $\uptau_{\textrm{MR}}$ which is contributed by the spin-lattice relaxation with the characteristic time $\uptau_{\textrm{SLR}}$ and the spin diffusion. The relaxation times are equal ($\uptau_{\textrm{MR}}=\uptau_{\textrm{SLR}}$) for disordered DMS alloys and/or in the case when the spatial inhomogeneity in the Mn heating does not contribute to the magnetization relaxation. These conditions are suggested to apply implicitly or explicitly to most of the experiments, but this is not the case of our study, since the structures have been designed to evolve the spin diffusion contribution.

Measurements were performed at low temperature \linebreak $T=1.8$~K for samples immersed in pumped liquid helium and in external magnetic fields $B$ up to 10~T. The magnetic field generated by a superconducting split-coil solenoid was applied parallel to the structure growth axis ($z$-axis). The photoluminescence signal was detected for right-hand ($\sigma^+$) circular polarization, which is the strongest component in applied external magnetic fields.

A pump-probe technique with time-resolved detection of the PL signal excited by the probe laser pulses has been used. A pulsed YAG laser with a photon energy of 3.49~eV (wavelength of 355~nm) was used as pump to provide heating of the Mn spin system. It generated pulses with 5~ns duration at a repetition frequency ranging from 100~Hz up to 5~kHz. Excitation power densities up to $ \gtrsim 200$~kW/cm$^2$ were reached. A semiconductor laser (photon energy 3.06~eV, wavelength 405~nm) with external amplitude modulation operated as the probe laser. Laser pulses with a duration from 0.5 up to 100~$\upmu$s were synchronized with the pump pulses. The time delay between the pump and probe pulses was varied by a digital delay generator in the range from 0 up to 10~ms. The excitation density of the probe laser has been kept below 0.1~W/cm$^2$ in order to minimize its contribution to the heating of the Mn spin system. The time-resolved PL spectra were recorded by means of a gated charge-coupled-device (CCD) camera with a temporal resolution of $\lesssim 2$~ns. Avoidance of a spatially inhomogeneous excitation has been realized by overlapped pump and probe excitation spots with diameters of $\sim 500~\upmu$m and by collection of the PL signal from their central part with a diameter of $\sim 50~\upmu$m.

\section{Magnetization dynamics in type-II DMS heterostructures}

The DMS heterostructures Zn$_{1-x}$Mn$_x$Se/Be$_{1-y}$Mn$_y$Te have a type-II band alignment with a potential energy minimum for conduction band electrons in the Zn$_{1-x}$Mn$_x$Se layers and for valence band holes in the Be$_{1-y}$Mn$_y$Te layers. The large difference in the band gaps, which are about 2.8~eV for Zn$_{1-x}$Mn$_x$Se and 4.5~eV for Be$_{1-y}$Mn$_y$Te, and huge band offsets of $\gtrsim 2$ and $\sim 1$~eV for conduction and valence bands, respectively, prevent the penetration of the electron and hole wave functions into the neighboring layers. Details of the growth, band structure, and optical properties of these structures can be found in previous publications~\cite{Yak01,Yak02}.

We have studied three MQW samples with a type-II band alignment: sample \#1 with Zn$_{0.99}$Mn$_{0.01}$Se/BeTe (20/10 nm), sample \#2 with Zn$_{0.99}$Mn$_{0.01}$Se/Be$_{0.93}$Mn$_{0.07}$Te (20/10 nm), and sample \#3 with Zn$_{0.99}$Mn$_{0.01}$Se/Be$_{0.93}$Mn$_{0.07}$Te (10/5 nm). Each sample has ten periods with a respective layer thickness of 20, 10 or 5~nm. Sample \#1 does not contain Mn in the BeTe layers. It serves as a reference structure to the sample \#2, which only differs by the presence of Mn in the Be$_{0.93}$Mn$_{0.07}$Te layers. The main difference between the samples \#2 and \#3 is the thickness of the Zn$_{0.99}$Mn$_{0.01}$Se layers of 20 and 10~nm, respectively. This results in different distances for spin diffusion from the center of these layers to the interfaces with  Be$_{0.93}$Mn$_{0.07}$Te.

\begin{figure}[t]%
\includegraphics*[width=\linewidth]{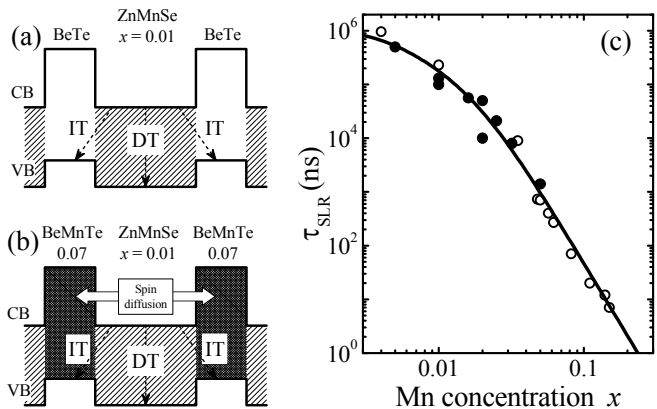}
\caption{%
  (a) and (b) Schematic band diagrams of Zn$_{1-x}$Mn$_x$Se/Be$_{1-y}$Mn$_y$Te quantum wells with a type-II band alignment. Dashed arrows show spatially direct (DT) and indirect (IT) optical transitions. Presence of the Mn ions in the Be$_{1-y}$Mn$_y$Te layers, see panel (b), opens a new channel for magnetization relaxation in the Zn$_{0.99}$Mn$_{0.01}$Se layer via spin diffusion into the Be$_{1-y}$Mn$_y$Te layers, see text. (c) Spin-lattice relaxation time as a function of the Mn content in II-VI DMS structures based on Cd$_{1-x}$Mn$_x$Te (closed circles) and Zn$_{1-x}$Mn$_x$Se (open circles) (Ref.~\onlinecite{Kneip06b}); the solid line represents the interpolation of this dependence.} \label{f1}
\end{figure}

\begin{figure}[t]%
\includegraphics*[width=0.9\linewidth]{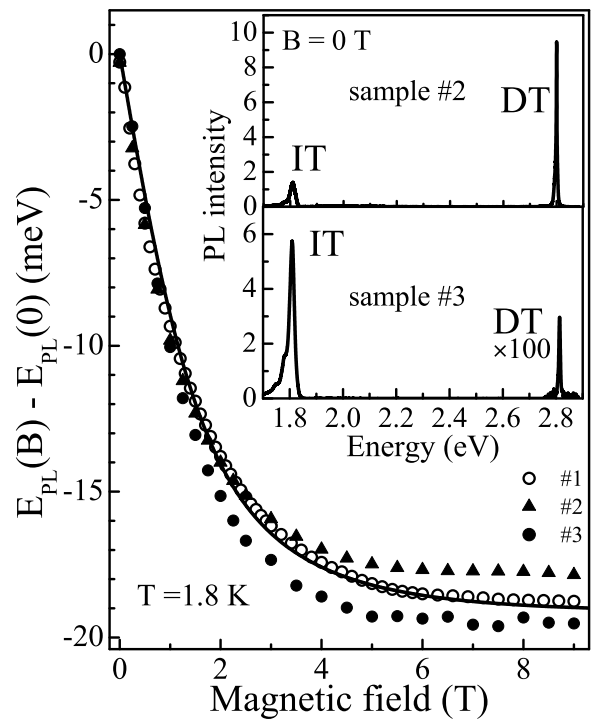}
\caption{%
Giant Zeeman shift of the spatially direct PL line in different Zn$_{1-x}$Mn$_x$Se/Be$_{1-y}$Mn$_y$Te QWs. Experimental data are shown by symbols, the solid line is a fit for the sample $\#1$ with parameters given in the text. In the insets the photoluminescence spectra of the samples \#2 and \#3 are shown.} \label{f2}
\end{figure}

In Fig.~\ref{f1}(a,b) the band diagrams for the studied type-II structures are shown schematically. The spatially direct (DT) and indirect (IT) optical transitions are marked by arrows. In the photoluminescence spectra they appear as lines at $\sim 2.8$ and $\sim 1.8$~eV, respectively, as can be seen in the insets of  Fig.~\ref{f2}. In our experimental conditions the photons, generated by the pump as well as the probe laser, are absorbed by the Zn$_{1-x}$Mn$_x$Se layers only. Hence, the indirect emission line is attributed to the holes scattered from the Zn$_{1-x}$Mn$_x$Se into the Be$_{1-y}$Mn$_y$Te layers after the photogeneration. This hole relaxation process is fast and depends strongly on the width of the Zn$_{1-x}$Mn$_x$Se layer~\cite{jetp2006,jetp2008}. As a result, the re\-lative intensities of the direct and indirect emission lines in the samples \#2 and \#3 (pay attention to the multiplication factor for the DT line of the sample \#3) differ considerably. Since we are interested in the magnetization dynamics of the Zn$_{1-x}$Mn$_x$Se layers, we will focus on the giant Zeeman shift of the DT line in the following.

Due to the strong exchange interaction of the free carriers with the localized spins of the Mn$^{2+}$ ions the exciton transitions in DMS show a giant Zeeman shift in external magnetic fields. This shift equals to one half of the giant Zeeman splitting of the exciton commonly described by the modified Brillouin function, for details see, e.g., Eqs. (1) and (2) in Ref.~[\onlinecite{Kneip06b}]. The experimental results for the giant Zeeman shift in the three studied samples are depicted in Fig.~\ref{f2}. All three samples show very similar shifts saturating at the level of $\sim$18--20~meV, which confirms that the static magnetic properties of the Zn$_{0.99}$Mn$_{0.01}$Se layers are not affected by the presence or absence of the Mn ions in adjacent layers. This is a predictable behavior, as the Mn-carrier exchange interaction is short-range and the carrier wave functions are strongly localized in the Zn$_{0.99}$Mn$_{0.01}$Se layers\cite{jetp2008,pss2000}. The solid line in Fig.~\ref{f2} is a fit by the modified Brillouin function of the experimental data for the sample \#1. It has been done with the following parameters: $S_{\textrm{eff}}=2.26$, $T_0=1$~K and $x=0.011$. Similar fits performed for the samples \#2 and \#3 show that the Mn concentrations in the Zn$_{1-x}$Mn$_x$Se layers of all three samples differ by less than 8\%. Therefore, their dynamical characteristics can be directly compared with each other.

\begin{figure}[t]%
\includegraphics*[width=0.9\linewidth]{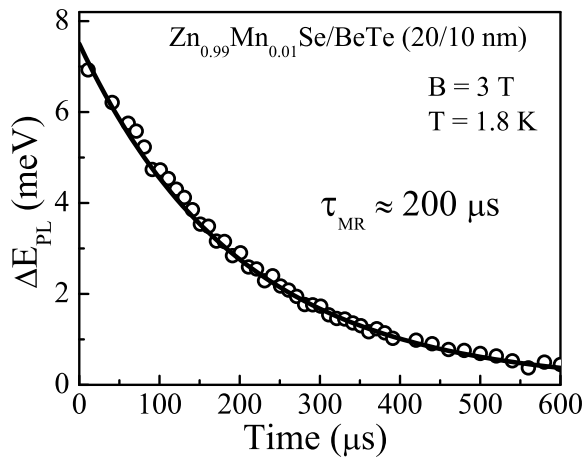}
\caption{%
Temporal evolution of the changes in the spectral position of the PL line $\Delta E_{\text {PL}}$  in Zn$_{0.99}$Mn$_{0.01}$Se/BeTe (20/10 nm) MQW, sample \#1.} \label{f3}
\end{figure}

\begin{figure}[t]%
\includegraphics*[width=0.9\linewidth]{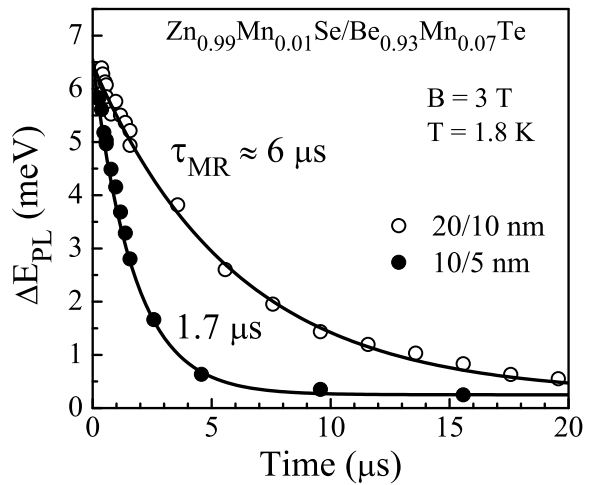}
\caption{%
Temporal evolutions of the changes in the spectral position of the PL line $\Delta E_{\text {PL}}$ in two Zn$_{0.99}$Mn$_{0.01}$Se/Be$_{0.93}$Mn$_{0.07}$Te structures with different layer thicknesses:  20/10 nm (\#2 -- open circles) and 10/5 nm (\#3 -- closed circles). } \label{f4}
\end{figure}

The dynamics of the magnetization relaxation after the pump pulse heating are illustrated for all three samples in Figs.~\ref{f3} and \ref{f4}. In the sample \#1 the relaxation time $\uptau_{\textrm{MR}}\approx200~\upmu$s coincides well with the SLR time known for Zn$_{0.99}$Mn$_{0.01}$Se QWs~\cite{Kneip06b}. Indeed, for this sample no contribution of spin diffusion to the magnetization relaxation is expected, since only the Zn$_{0.99}$Mn$_{0.01}$Se layers contain Mn ions.

In comparison with the sample \#1 a very strong reduction of $\uptau_{\textrm{MR}}$ down to 6~$\upmu$s is found in the sample \#2. Obviously, this is due to the spin diffusion from the Zn$_{0.99}$Mn$_{0.01}$Se layers to the Be$_{0.93}$Mn$_{0.07}$Te layers, where the spin-lattice relaxation time is expected to be very fast ($\lesssim 200$~ns), see Fig.~\ref{f1}(c). Note, that very close results have been observed in similar samples in experiments, where pulses of nonequilibrium phonons have been used for the heating of the Mn spin system~\cite{Scherbakov05}.

A further shortening of $\uptau_{\textrm{MR}}$ down to 1.7~$\upmu$s in the sample \#3 by a factor of about 4 (in comparison with the sample \#2) originates from a twice smaller distance for spin diffusion between the centers of the Zn$_{0.99}$Mn$_{0.01}$Se layers and the interfaces to Be$_{0.93}$Mn$_{0.07}$Te. We conclude from these experimental data that the spin diffusion in Zn$_{0.99}$Mn$_{0.01}$Se can provide a spin transfer over distances exceeding 10~nm. We will show in the following Section IV that the results of Figs.~\ref{f3} and \ref{f4} are sufficient for the evaluation of the spin diffusion coefficient. It makes heteromagnetic structures very suitable for quantitative studies of the spin diffusion in DMS.

\section{Numerical simulations of magnetization dynamics}

The magnetic relaxation time $\uptau_{\textrm{MR}}$, extracted from the dynamical shift of the PL line, can be treated as the spin-lattice relaxation time $\uptau_{\textrm{SLR}}$ only in case of uniformly distributed disordered magnetic ions. As it has been discussed in Sec.~III, the spin diffusion in the structures with a nonuniform Mn distribution may strongly accele\-rate the magnetic relaxation. Generally, when the Mn concentration $x(z)$ is a function of the growth direction $z$, the magnetization $M(z)$ and the Mn spin temperature $T_{\textrm{Mn}}(z)$ are also dependent on $z$ during the magnetization relaxation process. Hence, the spectral position of the PL line, $E_{\textrm{PL}}$, is determined by the magnetization in the regions averaged by the wavefunctions of the carriers involved in the recombination. In this case the information about the temporal changes of the spatial profile of $T_{\textrm{Mn}}$ and, thus, about the local magnetization is very important for the adequate description of the experimentally obtained dynamics of $\Delta E_{\textrm{PL}}(t)$. We have developed a numerical model suitable for a direct comparison with experimental results in different samples.

\subsection{Concentration dependence of SLR dynamics}

The experimental results for the SLR dynamics of the Mn spin system are summarized in Fig.~\ref{f1}(c), where the SLR time is plotted as a function of the Mn concentration. The open circles are data for Zn$_{1-x}$Mn$_x$Se-based structures, the data for Cd$_{1-x}$Mn$_x$Te are illustrated by the closed symbols~\cite{Kneip06b}. The SLR times cover five orders of magnitude from $10^{-3}$ down to $10^{-8}$ seconds as the Mn concentration varies from 0.004 up to 0.11. Such a strong dependence evidences that Mn-Mn interactions play a key role in the SLR dynamics. The results for both materials closely coincide demonstrating that the SLR of the Mn$^{2+}$ ions is rather insensitive to the ion host material.

We turn now to the physical mechanisms responsible for the strong dependence of the SLR time $\uptau_{\textrm{SLR}}(x)$ on the Mn concentration. It is known that an isolated Mn$^{2+}$ ion does not couple with the lattice in a perfect II-VI semiconductor crystal~\cite{Abr70} and, therefore, the SLR time should be extremely long. This expectation can be explained by the following reasons: The electric field of the phonons does not act on the magnetic moments of the $3d$ electrons and also the magnetic field, induced by the temporal variation of the electric field, is relativistically small. In principle, gradients of the electric field may interact with the quadrupolar moment of the magnetic ion which, in turn, interacts with the spin. But for the Mn$^{2+}$ ions the magnetic $d$-shell is half-filled and its quadrupolar moment is equal to zero. In this case the dominating mechanism of the spin-phonon interaction is caused by the phonon modulation of the spin-spin interaction between neighboring Mn ions. The phonons modulate the distance between these ions, hence, the local exchange field of an ion fluctuates. It is known as the Waller mechanism \cite{Abr70}, which is obviously strongly dependent on the concentration of magnetic ions.
Only the anisotropic exchange interaction is relevant because its operator does not commute with the Zeeman one.

The spin-spin interactions for the Mn$^{2+}$ ions are provided by three mechanisms: the exchange interaction, the Dzyaloshinsky-Moriya interaction~\cite{Dzy58, Mor60} and the magneto-dipole interaction. All of them decrease with increasing distance between the Mn ions, but the magneto-dipole interaction has the longest range. Therefore, despite being relatively weak, it plays the dominant role for SLR dynamics in the limit of small Mn concentrations. The exchange and Dzyaloshinsky-Moriya interactions gain in importance at higher Mn contents, when the formation of Mn clusters is more probable. Both the number of Mn spins coupled in clusters as well as the typical cluster size progressively increase with growing Mn concentration~\cite{Lar86}. This is the reason for the strong dependence of the SLR time on the Mn concentration~\cite{Far96, Sca96, Wan92}. Within the framework of a simple model which accounts for the spin diffusion from Mn ions to Mn clusters, where the spin has an efficient relaxation, it can be shown that the SLR rate $1/\uptau_{\textrm{SLR}}$ has a strong dependence on the Mn content with a power law between $x^{3}$ and $x^{4}$~[\onlinecite{Suris}]. This is in good qualitative agreement with the experimental results of Fig.~\ref{f1}(c).

According to Fig.~\ref{f1}(c), one can conclude that for fixed external conditions the Mn concentration $x$ mainly determines the SLR dynamics. This establishes a relationship between the static and dynamic magnetization, e.g., between the giant Zeeman splitting value and the SLR time - both being dependent on the Mn content. This restricting relationship can be relaxed or even canceled by the growth of heterostructures with modulated Mn content and/or with modulation doping by donors or acceptors providing free carriers. In Section V few examples of such structures are regarded, where the magnetization dynamics has been considerably accelerated without changing the static magnetization.

\subsection{Model}

The magnetization $M(B,T_{\textrm{Mn}})$ in DMS is a function of the magnetic field strength and the Mn spin temperature~\cite{Furdyna88}:
\begin{equation}\label{eq3}
M(B,T_{\textrm{Mn}})= g \mu_{\textrm{B}} x N_0
S_{\textrm{eff}}(x)\textrm{B}_{5/2}\left[\frac{5g \mu_{\textrm{B}}
B}{2k_{\textrm{B}}(T_{\textrm{Mn}}+T_0(x))}\right].
\end{equation}
Here $N_0$ is the inverse unit-cell volume, B$_{5/2}$ is the modi\-fied Brillouin function, $g=2.01$ is $g$-factor of the Mn$^{2+}$ ions, $k_{\textrm{B}}$ is the Boltzmann constant and $\mu_{\textrm{B}}$ is the Bohr magneton. $S_{\textrm{eff}}(x)$ is the effective spin and $T_0(x)$ is the effective temperature. These two parameters account for the antiferromagnetic Mn-Mn interactions contributing to the overall magnetization~\cite{Gaj79,Kel02}. In the equilibrium condition the Mn spin temperature is equal to the temperature of the phonon bath: $T_{\textrm{Mn}}=T_{\textrm{L}}$.

The model should take into account the processes of spin diffusion and spin-lattice relaxation in the Mn spin system. It should finally describe the relaxation of the magnetization changes $|\Delta M(\Delta T_{\textrm{Mn}})|=|M(T_{\textrm{Mn}})-M(T_{\textrm{L}})|$ induced by an external pertubation in a constant external magnetic field. In this non-equilibrium situation the spin temperature difference is given by $\Delta T_{\textrm{Mn}}=T_{\textrm{Mn}}-T_{\textrm{L}}$. In DMS with a homogeneous Mn concentration $x$, which does not depend on spatial coordinates, the magnetization dynamics in the one-dimensional case can be described by the following equation:

\begin{equation}
\label{eq1}
\frac{\partial}{\partial t} |\Delta M| =
K_{\textrm{diff}}(x) \frac{\partial^2}{\partial z^2}|\Delta
M| -\frac{|\Delta M|}{\uptau_{\textrm{SLR}}(x)} \hspace{0.1in}.
\end{equation}
Here $K_{\textrm{diff}}(x)$ is the spin diffusion coefficient.  In the case of a dipole-dipole interaction between the Mn$^{2+}$ ion spins the dependence of $K_{\textrm{diff}}$ on the Mn concentration $x$ can be expressed as:
\begin{equation}\label{eq2}
K_{\textrm{diff}}(x) = K_0  x^{4/3},
\end{equation}
where $K_0$ is the spin diffusion constant, see Eq.~(3.25) in Ref.~[\onlinecite{Burshtein}].

In consideration of Eq.~(\ref{eq3}), the Eq.~(\ref{eq1}) can be written in terms of $\Delta T_{\textrm{Mn}}$:
\begin{equation}\label{eq4}
\frac{\partial}{\partial t}\Delta T_{\textrm{Mn}} =
K_{\textrm{diff}}(x) \frac{\partial^2}{\partial
z^2}\Delta T_{\textrm{Mn}} -\frac{|\Delta
M|}{\uptau_{\textrm{SLR}}(x)} \left(\frac{dM(T_{\textrm{Mn}})}{dT_{\textrm{Mn}}}\right)^{-1}.
\end{equation}
In the case of small deviations from the equilibrium state and/or if $M$ is a linear function of $T_{\textrm{Mn}}$, the last term becomes equal to $\Delta T_{\textrm{Mn}}/\uptau_{\textrm{SLR}}$. It should be noted that for the ranges of $B$ and $T_{\textrm{Mn}}$, regarded in the present publication, the relation between the magnetization changes $\Delta M$ and $\Delta T_{\textrm{Mn}}$ is close to linear. Therefore, we will suggest that the relaxation of $\Delta T_{\textrm{Mn}}$ is simply determined by $\uptau_{\textrm{SLR}}$.

The equation~(\ref{eq4}) can be generalized for a nonuniform Mn distribution along the growth direction $z$ of the hete\-rostructure $x(z)$:
\begin{equation}\label{eq5}
\frac{\partial}{\partial t} \Delta T_{\textrm{Mn}} =
\frac{\partial}{\partial
z}\left(K_{\textrm{diff}}(z)\frac{\partial}{\partial z}
\Delta T_{\textrm{Mn}}\right)
-\frac{\Delta T_{\textrm{Mn}}}{\uptau_{\textrm{SLR}}(z)}\hspace{0.1in}.
\end{equation}
In the model we suggest that the spin diffusion coefficient $K_{\textrm{diff}}(z)$ is determined by the local Mn concentration $x(z)$ according to Eq.~(\ref{eq2}). Additional assumption is that the SLR time is determined by the local Mn concentration $x(z)$. For the dependence $\uptau_{\textrm{SLR}}(x)$ we used the interpolation function shown in Fig.~\ref{f1}(c) by the solid line.

In order to compare the experimental and numerical results the calculations have been performed in two steps. Firstly, we have solved numerically the one-dimensional partial differential Eq.~(\ref{eq5}), which describes the dynamics of the spatial distribution of $T_{\textrm{Mn}}$ in the sample with a nonuniform concentration of magnetic ions, as an initial-boundary value problem. We have chosen the uniform in space distribution of $T_{\textrm{Mn}}$ as the initial condition. The solution gives us information about $T_{\textrm{Mn}}$ in different points in space and at different time moments. An example of such kind of calculation, which will be discussed in detail below, is shown in Fig.~\ref{f5}.

The second step comprises the numerical solution of the Schr\"{o}dinger equations for electrons and holes. The potential profiles for carriers in DMS depend on the external magnetic field strength and the Mn spin tempera\-ture due to the giant Zeeman effect. The corresponding energy shifts of conduction and valence bands are described by:
\begin{equation}\label{eq6}
    \Delta E_{e(h)}(B,T_{\textrm{Mn}}) = \frac{\upalpha_{e(h)}}{\mu_{\textrm{B}}g}  M(B,T_{\textrm{Mn}})  ,
\end{equation}
where $N_0 \upalpha_e$ and $N_0 \upalpha_h$ are the exchange constants for the conduction and valence band. In Zn$_{1-x}$Mn$_x$Se they are equal to 0.26 and -1.31~eV, and in Cd$_{1-x}$Mn$_x$Te they are equal to 0.22 and -0.88~eV, respectively~\cite{Die94}. The Mn spin temperature distribution $T_{\textrm{Mn}}(z,t)$, obtained as numerical solution of the Eq.~(\ref{eq5}), allows us to determine the exact potential profiles for electrons and holes in an external magnetic field at different times after the pulsed heating of the Mn spin system. For every moment of time we solved numeri\-cally the Schr\"{o}dinger equations to obtain the electron and hole energies. This, in turn, allows us to calculate the changes in the spectral position of the PL line and to compare it with the experimental data of $\Delta E_{\textrm{PL}}(t)$.

\subsection{Numerical results for type-II DMS heterostructures}

\begin{figure}[t]%
\includegraphics*[width=0.9\linewidth]{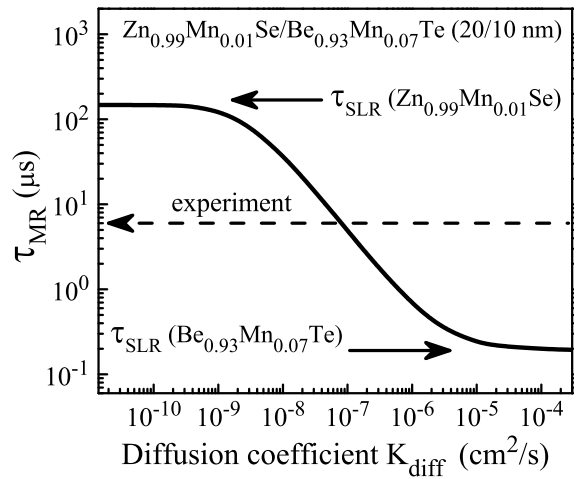}
\caption{%
Calculated dependence of the magnetic relaxation time $\uptau_{\textrm{MR}}$ on the diffusion coefficient $K_{\textrm{diff}}$ in the Zn$_{0.99}$Mn$_{0.01}$Se layer of the Zn$_{0.99}$Mn$_{0.01}$Se/Be$_{0.93}$Mn$_{0.07}$Te (20/10~nm) QW. In the limiting cases of small and large $K_{\textrm{diff}}$ the magnetic relaxation time is determined by $\uptau_{\textrm{SLR}}$ in Zn$_{0.99}$Mn$_{0.01}$Se and Be$_{0.93}$Mn$_{0.07}$Te, respectively. The dashed line shows the experimental value of $\uptau_{\textrm{MR}}\approx6$~$\upmu$s.} \label{f8}
\end{figure}

\begin{figure}[b]%
\includegraphics*[width=0.9\linewidth]{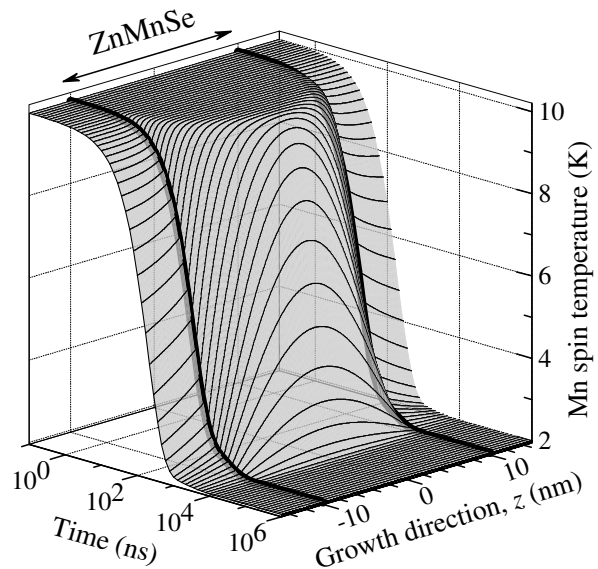}
\caption{Calculated temporal and spatial distribution of the Mn spin temperature in the Zn$_{0.99}$Mn$_{0.01}$Se/Be$_{0.93}$Mn$_{0.07}$Te QW (20/10 nm). Initially, $T_\textrm{Mn}=10$~K is uniform in space, the final temperature is $T_\textrm{Mn}=T_\textrm{L}=2$~K, $K_{\textrm{diff}} = 7\cdot 10^{-8}~\textrm{cm}^2/\textrm{s}$ in the Zn$_{0.99}$Mn$_{0.01}$Se layer. Thick solid lines show the temperature behavior close to the interfaces. } \label{f5}
\end{figure}

In this Section we present the results of numerical simulations of the magnetic relaxation dynamics in the type-II Zn$_{0.99}$Mn$_{0.01}$Se/Be$_{0.93}$Mn$_{0.07}$Te (20/10 nm) MQW (sample $\#$2). The experimental results for this structure are shown in Fig.~\ref{f4}. The Mn profile $x(z)$ in this structure, which we used to solve Eq.~(\ref{eq5}), is a step-like function with the values of 0.01 and 0.07 in the adjacent layers. The SLR times $\uptau_{\textrm{SLR}}$ in Zn$_{0.99}$Mn$_{0.01}$Se ($\sim 200~\upmu$s) and Be$_{0.93}$Mn$_{0.07}$Te ($\lesssim 200$~ns) differ by more than three orders of magnitude. In that case the Be$_{0.93}$Mn$_{0.07}$Te layers serve as a very efficient SLR channel for spin excitations which diffuse from the Zn$_{0.99}$Mn$_{0.01}$Se layers to them. The results of the numerical simulations become nearly independent on the spin diffusion inside the Be$_{0.93}$Mn$_{0.07}$Te layers and are predominantly controlled by $K_{\textrm{diff}}$ in the Zn$_{0.99}$Mn$_{0.01}$Se layers. Therefore, this type of heteromagnetic heterostructure is very reliable for an evaluation of $K_{\textrm{diff}}$.

\begin{figure}[bt]%
\includegraphics*[width=0.9\linewidth]{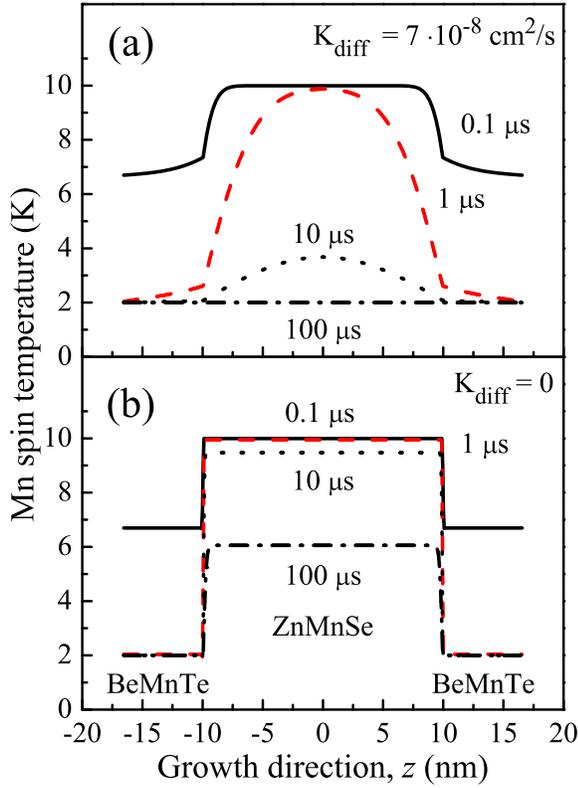}
\caption{%
(Color online) Calculated spatial distribution of the Mn spin tempera\-ture in the Zn$_{0.99}$Mn$_{0.01}$Se/Be$_{0.93}$Mn$_{0.07}$Te QW (20/10~nm) at different time delays after the pump pulse. Panels (a) and (b) correspond to different values of $K_{\textrm{diff}}$ in the Zn$_{0.99}$Mn$_{0.01}$Se layer. The initial and final temperatures are 10 and 2~K, respectively.} \label{f6}
\end{figure}

In Figure~\ref{f8} the calculated dependence of $\uptau_{\textrm{MR}}$ on $K_{\textrm{diff}}$ is shown by the thick solid line. The magnetic relaxation times for this dependence were used as characteristic times for the calculated $\Delta E_{\textrm{PL}}(t)$ dynamics. These times were taken at the point where the normalized $\Delta E_{\textrm{PL}}(t)$ is reduced to $1/e=0.37$ of its original value. Thus, the results in Fig.~\ref{f8} can be directly compared with the experiment. As expected, the extremal values of $K_{\textrm{diff}}$ $\rightarrow 0$ and $\rightarrow\infty$ reflect the identity $\uptau_{\textrm{MR}}=\uptau_{\textrm{SLR}}$ in the Zn$_{0.99}$Mn$_{0.01}$Se and Be$_{0.93}$Mn$_{0.07}$Te layers, respectively. The dashed arrow in  Fig.~\ref{f8} indicates the experimental value for $\uptau_{\textrm{MR}}\approx6~\upmu$s, measured for the sample \#2 (see Fig.~\ref{f4}). From its crossing point with the solid line we evaluate the spin diffusion coefficient in Zn$_{0.99}$Mn$_{0.01}$Se with rather high accuracy to $K_{\textrm{diff}} \approx (7\pm1.5)\cdot 10^{-8}~\textrm{cm}^2/\textrm{s}$. Then, according to Eq.~(\ref{eq2}) the diffusion constant is $K_0 \approx 3.5\cdot 10^{-5}~\textrm{cm}^2/\textrm{s}$. To the best of our knowledge, this is the first experimental evaluation of the spin diffusion parameters in DMS.

\begin{figure}[bt]%
\includegraphics*[width=0.9\linewidth]{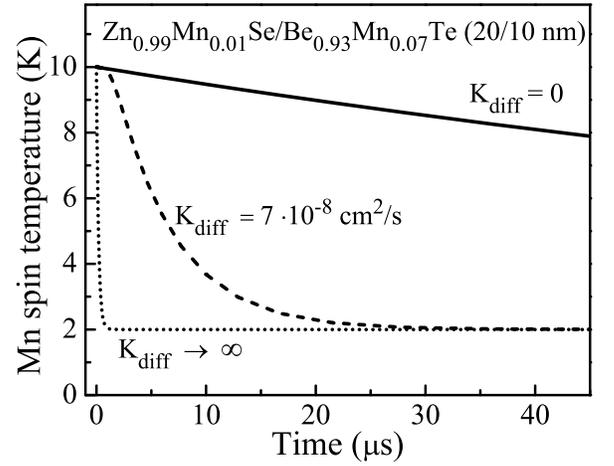}
\caption{%
Temporal evolution of the Mn spin temperature in the center of the Zn$_{0.99}$Mn$_{0.01}$Se layer calculated for different values of the diffusion coefficient $K_{\textrm{diff}}$. The initial and final temperatures are 10 and 2~K, respectively.} \label{f7}
\end{figure}

Let us make now a closer look at the calculated spatial and temporal profiles of $T_{\textrm{Mn}}$ in the heteromagnetic structure $\#$2 (see Fig.~\ref{f5}). The initial condition for this numeri\-cal problem is that the Mn spin system of the whole sample is heated homogeneously up to $T_{\textrm{Mn}} = 10$~K, which can be seen as a flat temperature level at short time delays. In our experiments such heating is provided by the interaction of the Mn ions with photogenerated carriers. The subsequent magnetic relaxation in the Mn spin system is provided by the SLR and by the spin diffusion. The latter is responsible for the spatially inhomogeneous cooling of the Mn spin system in the Zn$_{0.99}$Mn$_{0.01}$Se layer, which is clearly visible in Fig.~\ref{f5}. At time delays exceeding 20~$\upmu$s all Mn spins reach the equilibrium temperature of the phonon bath, which has been chosen to $T_{\textrm{L}}=2$~K.

For more detailed insight the spatial profiles of $T_{\textrm{Mn}}(z)$ at different time delays are shown in Fig.~\ref{f6}(a). In fact, these profiles are cross sectional areas of Fig.~\ref{f5} at fixed time delays ranging from 0.1 up to 100~$\upmu$s. In order to demonstrate the importance of the spin diffusion for the magnetic relaxation in this structure we present in Fig.~\ref{f6}(b) the calculated results for the case, where the spin diffusion is absent, i.e., for $K_{\textrm{diff}}=0$. It is consistent with an independent relaxation of $T_{\textrm{Mn}}$ in the adjacent layers with the corresponding SLR times. The temperature profiles in each layer are flat. Also, the relaxation of $T_{\textrm{Mn}}$ for, e.g., Mn ions located in the vicinity of the center of the Zn$_{0.99}$Mn$_{0.01}$Se layer is considerably slower than in the case of a finite diffusion coefficient, as shown in Fig.~\ref{f6}(a).

Figure~\ref{f7} demonstrates the dynamics of $T_{\textrm{Mn}}$ in the center of the Zn$_{0.99}$Mn$_{0.01}$Se layer which is strongly dependent on $K_{\textrm{diff}}$ in this layer. For $K_{\textrm{diff}}=0$ the spin diffusion is absent and the relaxation process is solely controlled by the SLR time in the Zn$_{0.99}$Mn$_{0.01}$Se layer, which approximately amounts to $200$~$\upmu$s. On the contrary, if the value of $K_{\textrm{diff}}\rightarrow\infty$, and, thus, the spin diffusion time does not restrict the magnetic relaxation time, the Mn spin temperature drops very fast to the bath value of $T_{\textrm{Mn}}=T_{\textrm{L}}=2$~K, being only controlled by the SLR time ($\lesssim 200$~ns) in the Be$_{0.93}$Mn$_{0.07}$Te layers. The calculated dynamics of $T_{\textrm{Mn}}$ with $K_{\textrm{diff}} =7\cdot 10^{-8}~\textrm{cm}^2/\textrm{s}$, which corresponds to the experimental situation of the sample $\#$2, is shown by the dashed line in Fig.~\ref{f7}.

\section{Magnetization dynamics in C\lowercase{d}T\lowercase{e}/C\lowercase{d}M\lowercase{n}T\lowercase{e}
structures engineered by digital growth technique}

The digital growth technique has achieved about atomic precision by using molecular-beam epitaxy. It has been implemented for GaAs/AlAs heterostructures, where during the growth the wide-band-gap AlAs layers with a certain thickness (frequently a submonolayer) have been introduced into narrow-band-gap GaAs layers at strictly predefined positions~\cite{Gos94}. This technique allows a highly variable engineering of the shape of the confining potential in quantum well structures. Also, by growing periodic structures with layer thicknesses of the order of a single monolayer a digital alloy (DA) can be fabricated. Its properties can significantly differ from the properties of the common disordered alloys, e.g., Al$_x$Ga$_{1-x}$As.

The digital growth technique has been successfully established for DMS heterostructures~\cite{Woj95,Cro95}. For the CdTe/Cd$_{1-x}$Mn$_x$Te material system digital alloys and parabolic quantum wells have been grown~\cite{Kneip06c,Woj96}. In these structures the band gap profile is controlled by the implementation of the Mn ions, and, therefore, the magnetic properties become also modulated. It has been shown experimentally that the partial smoothing of the digitally grown Mn profile considerably reduces the possible modifications of the static magnetic properties due to the Mn intermixing with neighboring CdTe layers~\cite{Gaj,Gaj1}. Typically, the latter are measured optically via the giant Zeeman splitting of the excitons. However, the magnetization dynamics can be changed drastically~\cite{Kneip06c}. Such a difference in the static and dynamical magnetization properties can be explained by different mechanisms. The static magnetization is mainly contributed by the paramagnetic Mn spins, while the magnetization dynamics critically depends on the concentration of the Mn clusters which serve as spin relaxation centers.

In this Section we analyze the Mn spin dynamics in two types of CdTe/Cd$_{1-x}$Mn$_x$Te heterostructures: digital alloys and parabolic QWs. We will demonstrate that the spin diffusion in the Mn spin system plays a key role in its dynamics, which is also simulated numerically.

\begin{figure}[tb]%
\includegraphics*[width=\linewidth]{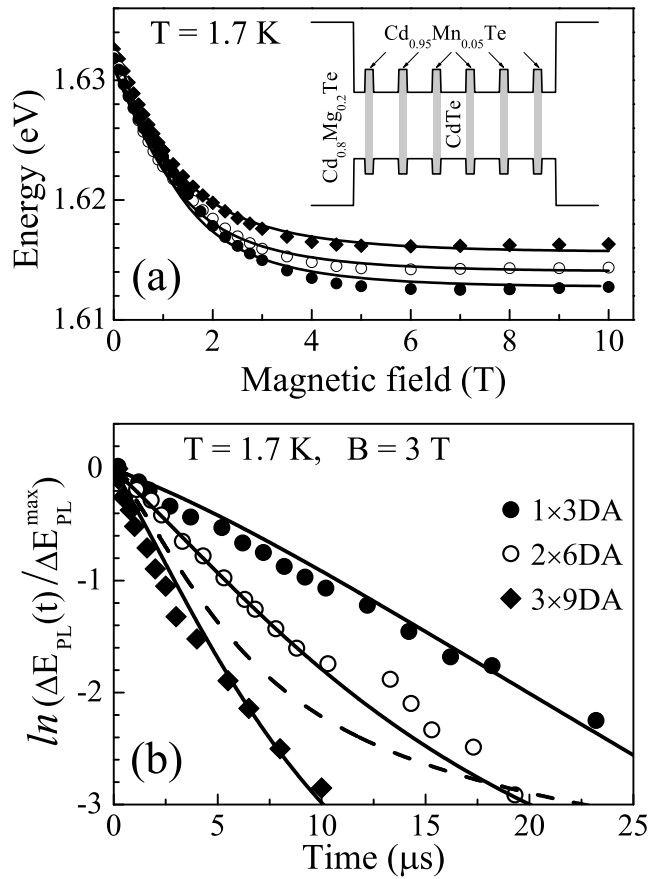}
\caption{
Giant Zeeman shift (a) and dynamical shift of the exciton PL lines $\Delta E_{\text {PL}}(t)$ during the magnetic relaxation (b) in three digital alloy samples. The sample design is shown schematically in the inset. The experimental data in both panels, shown by symbols, are taken from Ref.~[\onlinecite{Kneip06c}].  Solid lines are results of numerical calculations. For the modeling the values of the diffusion coefficient $K_{\textrm{diff}} = 3.5\cdot 10^{-5}x^{4/3} \textrm{[cm}^2/\textrm{s]}$  and Mn intermixing length $l_0=0.3$~nm were used. The dashed line in panel (b) represents the calculated results for the $2\times6$DA sample, neglecting spin diffusion ($K_{\textrm{diff}} = 0$).}
\label{f9}
\end{figure}

\subsection{CdTe/CdMnTe digital alloys}

Let us analyze the experimental results for three samples with CdTe/Cd$_{1-x}$Mn$_x$Te digital alloys reported in Ref.~[\onlinecite{Kneip06c}]. In these samples 23-nm-thick Cd$_{0.95}$Mn$_{0.05}$Te/CdTe DA layers (corresponding to about 70 monolayers) are confined between nonmagnetic Cd$_{0.8}$Mg$_{0.2}$Te barriers. According to the thickness of the magnetic and nonmagnetic layers, given in the monolayers, the structures are labeled in the following way: the first sample $1\times3$DA consists of 18 periods of 1ML/3ML, the second sample $2\times6$DA has 9 periods of 2ML/6ML, and the third sample $3\times9$DA has 6 periods of 3ML/9ML. The choice of the layer thickness provides the same average Mn concentration $x_{\textrm{DA}}\approx0.013$ in the three DA samples.

The experimental data for the excitonic giant Zeeman shifts and magnetic relaxation dynamics, measured in Ref.~[\onlinecite{Kneip06c}], are shown by the symbols in the Figs.~\ref{f9}(a) and \ref{f9}(b), respectively. One can see that the Zeeman shifts of all three samples follow closely each other with rather weak differences in the saturation levels at high magnetic fields. This reflects the fact that the static magnetization is mainly controlled by the average Mn concentration in the DA.

On the contrary, the magnetic relaxation dynamics of the digital alloys strongly depends on the details of the real (smoothed) Mn profiles. As one can see in Fig.~\ref{f9}(b), the magnetic relaxation times shorten from 9.5~$\upmu$s in $1\times3$DA to 2.5~$\upmu$s in $3\times9$DA. They considerably deviate from the relaxation time of 27~$\upmu$s, measured in the disordered Cd$_{0.985}$Mn$_{0.015}$Te reference alloy~\cite{Kneip06c}.

We have simulated numerically the magnetic relaxation dynamics in these digital alloys. For that purpose, we smoothed the nominal technological profile by implementing the Mn intermixing, i.e., we took the convolution of this profile with an exponential broadening function\cite{Gaj}:
\begin{equation}\label{eq7}
g(z) =\left\{
\begin{array}{ll}
\exp(z/l_0 -1),& z<l_0  \\
 0, & z\geqslant l_0 \quad .
  \end{array} \right.
\end{equation}
Then we solved the Eq.~(\ref{eq5}) by using the diffusion coefficient in form of Eq.~(\ref{eq2}) with $K_0=3.5\cdot10^{-5}$~cm$^2$/s, as it was estimated in Section~IV. We realized that the Mn intermixing length $l_0$ strongly controls the relative ratios of the three $\uptau_{\textrm{MR}}$ values in different DA samples, which offers a high precision for the evaluation of $l_0 \approx 0.3$~nm. It is important to stress here, that the magnetic relaxation in the digital alloys does not follow a single exponential law, however, it can be rather closely fitted, compare solid lines and symbols in  Fig.~\ref{f9}(b). It is also interesting that, despite the very small layer thicknesses in the DA, the consideration of the spin diffusion in the Mn spin system was important. The dashed line in Fig.~\ref{f9}(b) shows the calculated result for the $2\times6$DA sample in the absence of spin diffusion ($K_{\textrm{diff}} = 0$). It considerably deviates from the experimental data.

Using the Mn intermixing length $l_0=0.3$~nm, evaluated from the best fit of the magnetic relaxation dynamics, we have calculated the giant Zeeman shift and received good coincidence with the experimental data, as the solid lines and symbols in Fig.~\ref{f9}(a) indicate. This makes us confident that the used numerical approach properly accounts for the main factors which control the static and dynamic magnetization in the studied digital alloys.

\begin{figure}[hbt]%
\includegraphics*[width=\linewidth]{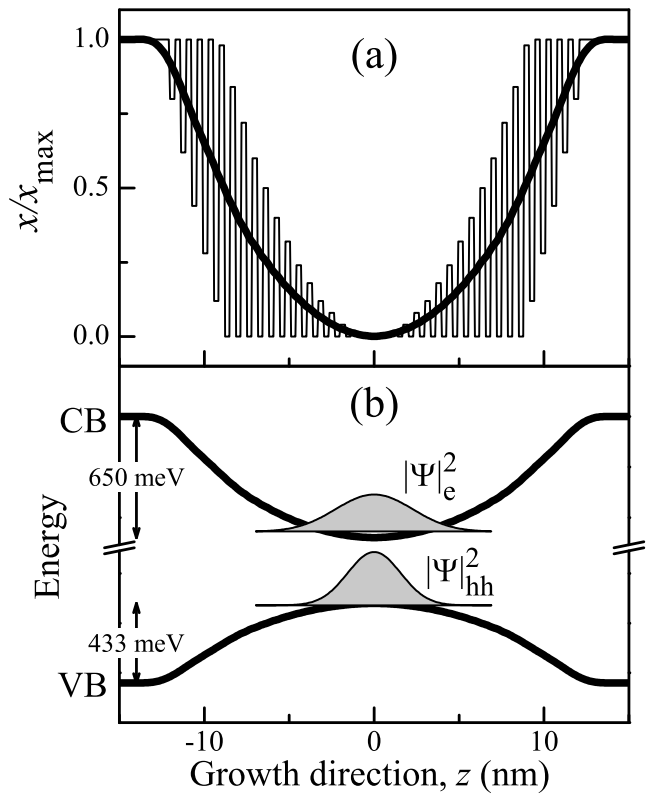}
\caption{ (a) The thin line represents the technological profile of the Mn concentration in a 25.4-nm-thick CdTe/Cd$_{1-x}$Mn$_x$Te parabolic QW. The effective averaged Mn concentration is given by the thick line (see text). (b) The effective conduction and valence band profiles for the parabolic QW with $x_{\textrm{max}}=0.68$. Gray areas show the squared moduli of the wave functions of electrons and heavy-holes in the ground states.} \label{f10}
\end{figure}

\subsection{CdTe/CdMnTe parabolic QWs}

The digital growth technology enables the fabrication of structures with a continuously varying Mn concentration and potential profile along the growth direction. One of the interesting examples of the technological capabilities is a parabolic CdTe/Cd$_{1-x}$Mn$_x$Te QW, where the confinement potential for the carriers in the QW has a parabolic dependence on the coordinate along the growth direction. Details of the growth and magneto-optical properties for such QWs can be found in Refs.~[\onlinecite{Woj96,Woj97,Oss98}].

We have measured the magnetic relaxation dynamics for two parabolic QWs with different maximal Mn contents $x_{\textrm{max}}$ of 0.13 and 0.68. Both QWs have a thickness of 25.4~nm, which corresponds to 82 monolayers (one monolayer of CdTe is equal to 0.324~nm). The Mn shutter was continuously opened for the growth of the barriers and pulsed during the growth of the parabolic QW. The growth of each QW was divided into 41 steps. In each step the Mn cell was opened for the time needed to obtain a given value of Mn concentration, more details can be found in Ref.~[\onlinecite{Woj96}]. The technological Mn profile averaged over each monolayer is shown by the thin solid line in Fig.~\ref{f10}(a). After its strong smoothing one can receive a parabolic profile represented by the thick line. In the real structures the Mn intermixing is not sufficiently strong to reach the monotonous Mn distribution shown by the solid line, it only smoothed the sharp boundaries in the nominal Mn distribution.

\begin{figure}[t]%
\includegraphics*[width=\linewidth]{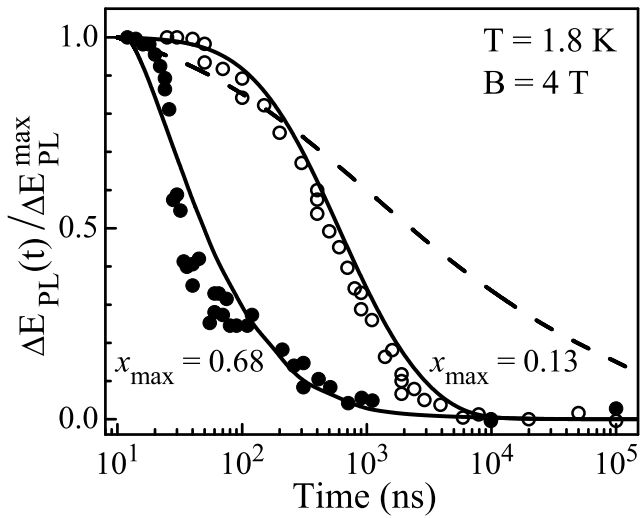}
\caption{ Magnetic relaxation dynamics in 25.4-nm-thick CdTe/Cd$_{1-x}$Mn$_x$Te parabolic quantum wells with different $x_{\textrm{max}}$.  Solid lines are calculation results with $K_{\textrm{diff}} = 3.5\cdot 10^{-5}x^{4/3} \textrm{[cm}^2/\textrm{s]}$  and $l_0=0.3$~nm. The dashed line shows the calculated results for the sample with $x_{\textrm{max}}=0.68$ without taking into account the spin diffusion ($K_{\textrm{diff}} = 0$).} \label{f11}
\end{figure}

The carrier wave functions, for both the electrons in the conduction band (CB) and the heavy-holes in the valence band (VB), penetrate through the digital layers. As a result, their quantum well confinement potentials can be well approximated by the parabolic functions shown in Fig.~\ref{f10}(b) by thick lines. The square modulus of each carrier wave function, calculated in the real potential with $x_{\textrm{max}}=0.68$ and by considering the Mn intermixing length $l_0=0.3$~nm, is shown by the gray area in Fig.~\ref{f10}(b). One can see that the electrons and holes at the lowest quantum confined energy levels are localized close to the center of the parabolic QW. The details on the energy spectrum and exciton giant Zeeman splitting in CdTe/Cd$_{1-x}$Mn$_x$Te parabolic QWs are given in Refs.~[\onlinecite{Woj97,Oss98}].

The experimental results on the magnetic relaxation dynamics, measured from the dynamical shift of the exciton PL line $\Delta E_{\text {PL}}$, are shown by the symbols in Fig.~\ref{f11}. The characteristic times $\uptau_{\textrm{MR}}$ for the samples with $x_{\textrm{max}}=0.13$ and 0.68 differ considerably from each other, being $\sim 1$~$\upmu$s and $\sim 100$~ns, respectively. It can be explained by the role of the Mn clustering, which controls the $\uptau_{\textrm{SLR}}$, and by the Mn spin diffusion.

One can expect that in these relatively thick structures with a gradual variation of the magnetic properties from the QW center to its edges the spin diffusion in the Mn spin system should play an important role for the magnetic relaxation. In fact, its significance for the sample with $x_{\textrm{max}}=0.68$ is pointed out by comparing the results of the numerical calculations with (the solid line in Fig.~\ref{f11}) and without (the dashed line) regarding the spin diffusion. The very long relaxation tail of the dashed line dependence (note the logarithmic time scale) is due to the long SLR times in the region of low Mn concentration near by the QW center. This region is important for the dynamics of the exciton giant Zeeman shift, as the electron and hole wave functions are primarily concentrated on the QW center, see Fig.~\ref{f10}(b).

Furthermore, the numerical results, presented in Fig.~\ref{f11}, do coincide with the experimental data for the parabolic QWs with very different $x_{\textrm{max}}=0.68$ and 0.13 without changing the calculation parameters ($K_{\textrm{diff}} = 3.5\cdot 10^{-5}x^{4/3} \textrm{[cm}^2/\textrm{s]}$, $l_0=0.3$~nm). This fact is very promising that the magnetic relaxation dynamics in such complicated structures can be well described within the frame of a relatively simple model and without fitting parameters. It also makes us confident that the suggested functional dependence of the diffusion coefficient on the Mn content, given by Eq.~(\ref{eq2}), is well suitable for the modeling of the magnetization dynamics.

\section{Discussion and Conclusions}

Note, that even in bulk disordered DMS samples with a homogeneous distribution of Mn ions the spin diffusion can play a significant role for the magnetic relaxation dynamics. This fact can be attributed to at least two reasons: (i) Spatially inhomogeneous heating of the Mn spin systems, e.g., by a tightly focused laser beam. Based on typical parameters for the studied materials, we estimate that the scale of such an inhomogeneity should be smaller than 1~$\upmu$m. (ii) Intrinsic instabilities of a highly excited Mn spin system, which may lead to the formation of domains with  "hot" and "cold" spin temperatures, whose size and dynamics will be governed by the spin diffusion~\cite{Tyazh,Tyazh1,Vladimirova}. These conclusions are equally true for heterostructures containing layers with only one Mn concentration, e.g., DMS quantum wells confined by nonmagnetic barriers.

It is known, that the SLR dynamics of the Mn ions is strongly accelerated by an increase of the lattice temperature \cite{Sca96, Sch00} and also becomes faster in stronger external magnetic fields \cite{Str92, Sch00}. Clearly, these effects would modify the magnetic relaxation dynamics in heteromagnetic heterostructures, which remarkably depends on the absolute values of $\uptau_{\textrm{SLR}}$.

In addition to that, the spin diffusion can be a dominating factor for the magnetic relaxation dynamics in epitaxially grown DMS quantum dots. Such quantum dots are typically grown on the top of the (about one monolayer thick) wetting layer, whose effective Mn concentration (due to the very small thickness, see discussion in the Section for digital alloys) is considerably smaller than in the quantum dots. Experimental data available for such dots show a strongly nonexponential magnetic relaxation, which has been tentatively explained by the spin diffusion between the Mn spin systems of the quantum dots and the wetting layer~\cite{Scherbakov04,Aki06a}. DMS quantum dots have a faster SLR compared to the wetting layer, but they are characterized by a very limited capacity of the spin reservoir. In this case, the spin diffusion would control the heating of the Mn spin system in contrast to the cooling scenarios, we have considered in the present paper.

In conclusion, we have performed detailed experimental and theoretical investigations of the magnetic relaxation dynamics in different types of DMS quantum wells. We have shown that the spin diffusion in the Mn spin system plays an important role in this dynamics, which is especially evident for the structures with a nonuniform Mn distribution. The diffusion coefficient $K_{\textrm{diff}} = 7\cdot 10^{-8}~\textrm{cm}^2/\textrm{s}$ has been evaluated for Zn$_{0.99}$Mn$_{0.01}$Se. The results for the parabolic QWs suggest that the functional dependence of the diffusion coefficient on the Mn content in terms of $K_{\textrm{diff}} \approx 3.5\cdot 10^{-5}x^{4/3}$~[cm$^2$/s] received from the theoretical approach of Ref.~[\onlinecite{Burshtein}] is suitable for the modeling in a wide range of Mn concentrations.

\begin{acknowledgements}
We greatly acknowledge the encouraging discussions with R.A.~Suris and D.~Dunker for the help with experi\-ment. The work has been supported by the Deutsche Forschungsgemeinschaft, the EU Seventh Framework Programme (Grant No. 237252, Spin-optronics) and by the Russian Foundation of Basic Research (Grants \#08-02-01302 and \#10-02-00549).
\end{acknowledgements}

\end{document}